# Design of Optical Metamaterial Mirror with Metallic Nanoparticles for Broadband Light Absorption in Graphene Optoelectronic Devices

*Seungwoo Lee[1*], Juyoung Kim[2]*

[1]SKKU Advanced Institute of Nanotechnology (SAINT) & School of Chemical Engineering, Sungkyunkwan University (SKKU), Suwon 440-746, Republic of Korea

[2]Department of Neurobiology, Stanford University School of Medicine, Stanford, CA 94305, USA

*Email: seungwoo@skku.edu

**Abstract**

A general metallic mirror (i.e., a flat metallic surface) has been a popular optical component that can contribute broadband light absorption to thin-film optoelectronic devices; nonetheless, such electric mirror with a reversal of reflection phase inevitably causes the problem of minimized electric field near at the mirror surface (maximized electric field at one quarter of wavelength from mirror). This problem becomes more elucidated, when the deep-subwavelength-scaled two-dimensional (2D) material (e.g., graphene and molybdenum disulfide) is implemented into optoelectronic device as an active channel layer. The purpose of this work was to conceive the idea for using a charge storage layer (spherical Au nanoparticles (AuNPs), embedded into dielectric matrix) of the floating-gate graphene photodetector as a magnetic mirror, which allows the device to harness the increase in broadband light absorption. In particular, we systematically examined whether the versatile assembly of spherical AuNP monolayer within a dielectric matrix (i.e., optical metamaterial mirror), which should be designed to be placed right below the graphene channel layer for floating-gate device, can be indeed treated as the effective magnetic mirror. In addition to being capable of the enhancement of broadband light absorption, versatile access to various structural motifs of AuNPs benefitting from recent advances in chemical synthesis promises compelling opportunities for sophisticated engineering of optical metamaterial mirror. High amenability of the AuNP assembly with the semiconductor-related procedures may make this strategy widely applicable to various thin film optoelectronic devices. Our study thereby illustrates advantages in advancing the design of mirror for rational engineering of light-matter interaction within deep-subwavelength-scaled optoelectronic devices.

**1. Introduction**

In thin film optoelectronic technologies, for example, photodetectors and solar cells, there has usually been a trade-off between efficient photon absorption and the enhancement of electrical performance (increase in the efficiency of charge/carrier collection and photocurrent) [1−8]. The reduction in the thickness of active film (i.e., channel layer or light absorption layer) represents the increase in the efficiency of charge/carrier collection, but it is inherently accompanied by sacrificing light absorption efficiency. More importantly, this trade-off problem becomes more elucidated, when optoelectronic devices mainly rely on the deep-subwavelength-scaled two-dimensional (2D) materials such as graphene and molybdenum disulfide ($MoS_2$) for benefiting from their exotic electronic properties (e.g., high charge carrier mobility even at room temperature) [9−14]. Even if the unique capabilities of 2D materials together with their high amenability to the currently available semiconductor processes promises compelling opportunities in advancing various optoelectronic devices [9-14]; nonetheless, the inherently low optical cross-section of 2D materials remains as challenges for their practical applications to optoelectronic devices [12].

Meanwhile, the light trapping technologies have envisioned wide range of optoelectronic devices over the past three decades, from ray optics-based statistical strategy (e.g., Yablonovitch limit) to wave optical near-field manipulation (e.g., plasmonic nanoantennas) [15−24]. Even more exciting is various semiconductor/metal-relevant nanotechnologies' potential (e.g., electron beam/focused ion beam lithography, self-assembly, and synthesis of nanowires) for enabling sophisticated engineering of light trapping with exquisite control over geometries and shapes of 2D/3D nanoscaled meso-structures [25,26]; consequently, such light trapping technologies can further advance the aims of 2D materials-based optoelectronic devices by expanding the range of accessible optical cross-section [27−33]. However, much of the development of light trapping technologies especially for 2D material-based deep-subwavelength-scaled optoelectronic devices has been mainly to regard the use of plasmonic nanoantennas rather than other photonic structural motifs [27−33]. Although this method can address some of the aforementioned optical problems of 2D materials, the requirements for the direct incorporation of antennas into 2D materials place severe restrictions on the available design space of optoelectronic devices. Also, the broadband enhancement of light absorption has been a challenging with the use of such plasmonic nanoantennas. Thus, other optical structural motifs need to be further investigated in order to contain a rich diversity of the deep-subwavelength-scaled optoelectronic device architecture with the rational light trapping design.

Metallic backplane mirror, which itself can act as a contact or gate electrode, has proven to be a versatile component for light trapping in the thin optoelectronic devices, as allowing us broadband photon recycling even with simple structural architecture (just flat film) in stark contrast to plasmonic nanoantennas [34−38]; but, translation of a flat metallic mirror particularly into the deep-subwavelength-scaled optoelectronic devices still requires optical engineer to first address the problem of significantly reduced optical intensity within one-quarter of wavelength, which is generally observed in low surface impedance metallic mirror (i.e., electric mirror) [38]. In other words, the phase of the light reflecting off a mirror (reflection phase) should be quantitatively controlled with a high flexibility, while simultaneously being compatible with optoelectronic device architectures. Very recently, Esfandyarpour et al., has promoted this issue by using electron-beam lithographic texturing of metallic surface [38], but the problems of difficulty in large-area fabrication together with relatively low compatibility with wide range of 2D material optoelectronic device architecture have stymied its practical applications.

Herein, we theoretically propose the concept of multifunctional magnetic mirror, in which a versatile assembly of metallic nanoparticle monolayer can achieve the high impedance and the minimized phase reversal, while fully taking advantage of its nonvolatile memory functionality. Also, the ability to precisely tune the structural features of primitives (i.e., AuNPs) makes this direction highly flexible especially in terms of absorption control within the deep-subwavelength-scaled, 2D material optoelectronic devices. Finally, the assembled monolayer of metallic nanoparticles can be highly amenable to the fabrication process and architecture of the deep-subwavelength-scaled, 2D material optoelectronic device.

## 2. Optical properties of graphene optoelectronic devices

*2.1 Device architectures and numerical calculations*

To access whether metallic nanoparticle monolayer-enabled metamaterial mirror can be indeed effective in terms of light trapping in the deep-subwavelength-scaled optoelectronic device, we chose as a model system the pentacene-graphene nano floating-gate transistor memory (NFGTM), which we had identified as a non-volatile photodetector with high optical cross-section and quantum efficiency (schematic for the device architecture is presented in Fig. 1a) [39]. Our device consists of plastic substrate (polyethylene naphthalate, PEN), a 300 nm thick indium tin oxide (ITO) gate electrode, 30 nm thick aluminum oxide ($Al_2O_3$) blocking dielectric layer (i.e., 30 % oxide and 70 % aluminum), 160 nm thick cross-linked poly-4-vinylphenol (cPVP) dielectric matrix of AuNP monolayer (i.e., the layer of optical metamaterial mirror with the functionality of charge storage), and 25 nm thick pentacene-graphene hybrid active channel layer (deep-subwavelength-scaled). Especially, in order to induce the nonvolatile photonic memory functionality via floating-gate, the layer of optical metamaterial mirror made of AuNPs should be placed right below the active channel layer.

Toward verification of optical properties of our devices, we carried out 3D full field electromagnetic simulation by finite-difference, time-domain (FDTD) method with the boundary conditions detailed in Fig. 1b (red colored bold box highlighting perfectly matched layer (PML) and periodic boundary conditions (PBC)). AuNPs were modeled to be encapsulated by 1 nm organic surfactant (poly(diallyl dimethyl ammonium chloride), polyDADMAC) [40,41]. In order to avoid unrealistic point contact, the interface between the hexagonally close-packed AuNPs (main structural motif of optical metamaterial mirror in the current work) was truncated by 1 nm [42]. Unit cell is also indicated in Fig. 1b (blue dotted box); due to the hexagonal geometry, the optical properties of this structure are independent on the polarization of normally incident light. For the calculation of the complex permittivity of Au, Drude-critical model was implemented into FDTD code [43]; the complex dielectric constants of other materials including PEN, ITO, polyDADMAC, and pentacene (thin film, which was assumed to be dried at 25 °C) were experimentally measured by ellipsometry. The Bruggeman effective medium approximation allows us to obtain the complex permittivity of $Al_2O_3$ [42]:

$$n_{al} \frac{(\varepsilon_{Al}-\varepsilon)}{(\varepsilon_{Al}+2\varepsilon)} + n_{oxide} \left(\frac{\varepsilon_{oxide}-\varepsilon}{\varepsilon_{oxide}+2\varepsilon}\right) = 0 \qquad (1)$$

where $n_{al}$ and $n_{oxide}$ are the volume ratios of Al and $Al_2O_3$, respectively; $\varepsilon$ is the permittivity of $Al/Al_2O_3$ composite. The both $\varepsilon_{Al}$ and $\varepsilon_{oxide}$ were obtained by a modified Drude model and empirical measurement, respectively. In order to calculate the effective complex permittivity of graphene, the optical conductivity, $\sigma(\omega) = \sigma_{intra}(\omega) + \sigma_{inter}(\omega)$, with respect to Fermi energy was obtained by means of Kubo formula [30]:

$$\sigma_{inter}(\omega) = i\frac{e^2\omega}{\pi}\int_\Delta^\infty d\epsilon \frac{(1+\frac{\Delta^2}{\epsilon^2})}{(2\epsilon)^2-(\hbar\omega+i\Gamma)2} \times [f(\epsilon-E_F)+f(\epsilon+E_F)] \qquad (2)$$

$$\sigma_{intra}(\omega) = \frac{e^2}{\pi\hbar^2}\frac{i}{\omega+i\tau^{-1}}\int_\Delta^\infty d\epsilon \left(1+\frac{\Delta^2}{\epsilon^2}\right) + [f(\epsilon-E_F)+f(\epsilon+E_F)] \qquad (3)$$

where $f(\epsilon-E_F)$ is the Fermi distribution function with Fermi energy ($E_F$), $\Gamma$ indicates the broadening of the interband transitions, $\tau$ corresponds to the momentum relaxation time (herein, 250 fs was employed) caused by carrier intraband scattering, and $\Delta$ is a half bandgap energy from the tight-binding Hamiltonian near $K$ points of the Brillouin zone.

*2.2 Optical properties of NFGTM*

As presented in Fig. 2a, graphene itself generally shows poor optical cross-section at visible frequency of interest (~ absorption of 2.3 %); the gating of graphene results in the reduction of broadband absorption, since the reflection at the surface of graphene becomes enhanced with the decrease in Fermi level (e.g., -1000 meV). Only recently, have we found that the inherently low optical cross-section and quantum efficiency of graphene can be effectively addressed merely by the coating of pentacene on it (see the change in optical absorption after coating of 25 nm thick pentacene, highlighted by bold orange line in Fig. 2a) [39]. Especially, the efficient generation of electron-hole pair within pentacene and subsequent transferring of the generated electron-hole pair to graphene via capacitive coupling was found to dramatically increase the photoresponsibility and photodetectivity of the graphene optoelectronic devices [39]. More importantly, the incorporation of the AuNP layer (individually dispersed AuNPs within cPVP) right below the pentacene-graphene hybrid layer, acting as a charge storage layer as well, can give rise to the plasmonic back scattering and the further enhancement of light absorption, while taking full advantage of non-volatile photonic memory (i.e., floating-gate system) [39]. Indeed, the absorption behavior of such pentacene-graphene hybrid layer (bold purple line in Fig. 2a) strongly relies on the plasmonic scattering behavior of individual AuNP (see back scattering spectrum of the disc-shaped AuNP with 5 nm height and 10 nm diameter, embedded in 20 nm thick cPVP dielectric matrix, as shown in Fig. 2b): at the localized plasmonic resonance wavelength of AuNPs, the absorption of the photodetector can be further increased (Fig. 2a). Herein, the shape of AuNP especially was designed to be disc as with our experimental result [39]. This success to the further increase in the photoresponsibility of the pentacene-graphene photodetector by means of plasmonic light scattering of AuNP charge storage layer, together with high compatibility of the AuNPs implementation with the fabrication procedure of pentacene-graphene photodetector (e.g., sputtering or spin coating-enabled self-assembly), amply inspires the current work on the rational design of a charge storage layer (i.e., AuNP monolayer) to be useful for optical metamaterial mirror, in that provides a way for recycling broadband light, maintaining nonvolatile photonic memory function, and learning some of their perspectives.

## 3. Design of optical metamaterial mirror

*3.1 Monolayer of hexagonally close-packed, spherical AuNPs as optical magnetic mirror*

We now outline how high surface impedance of spherical AuNP monolayer with hexagonally closed-packed geometry, which is useful for a charge storage, can be harnessed to realize optical magnetic mirror with a minimized phase reversal. Much like other periodically bumpy metallic surfaces [38, 44−47], close-packed spherical AuNP monolayer within dielectric cPVP matrix (metallo-dielectric photonic/plasmonic hybrid crystals) itself can efficiently make the path of surface current less straightforward, so as to effectively reduce optical conductivity ($\sigma$) of metal surface. This reduced $\sigma$ in turn results in the enhancement of $E_z$ at the surfaces of metallic structure. Particularly, in the case of metallo-dielectric photonic/plasmonic hybrid crystals working at optical frequency [38], both surface plasmon polaritons (SPPs) and photonic (PhC) modes can lead to the dramatic reduction of $\sigma$ (enhancement of $E_z$). As the impedance ($Z$) at the surface of metallic structures is given by the $E_z/H_y$ (following Ohm's law), the increase in $E_z$ at the surfaces, resulting from both SPPs and PhC modes, leads to the reduction of the reflection according to following complex reflection coefficient [38, 48]:

$$r = r_0 e^{i\varphi} = \frac{Z_{S1}\cos(\theta_i) - Z_{S2}\cos(\theta_t)}{Z_{S1}\cos(\theta_i) + Z_{S2}\cos(\theta_t)} \quad (4)$$

where $Z_{S1}$, $Z_{S2}$, $\varphi$, $\theta_i$, $\theta_t$, and $r_0$ denote the impedance at the metal surface, the wave impedance of the incident medium, reflection phase, incident angle, transmitted angle, and reflection amplitude, respectively. For the normal irradiation, second terms of the numerator and dominator of eqn. (4) become unity; then, the magnitude of reflection is simplified by $Z_{S1}$ (or $\sigma$) of the monolayer of close-packed spherical AuNPs ($r=(Z_{S1}-1)/(Z_{S1}+1)$). Thus, the increase in $Z_{S1}$ via SPPs or PhC modes allows us to minimize the phase reversal ($\varphi = \pi$).

*3.2 Properties of optical metamaterial mirror*

Fig. 3 presents the collective set of optical properties of metamaterial mirror layer (i.e., hexagonally close-packed, 150 nm sized spherical AuNP monolayer embedded into 160 nm thick cPVP matrix, which is a structural primitive in this work) including the dispersion relationship (band structure, obtained under the Bloch's boundary conditions), vertical electric field ($E_z$) distribution at the bandedge of $k_xa/2\pi$, impedance, and reflection. The spatial distribution of $E_z$ (Fig. 3b) reveals the nature of each different mode, presented in Fig. 3a, as follows. The $E_z$ of PhC mode 1 (left panel of Fig. 3b) is found to be mostly oriented along the horizontal direction, whereas the spatial distributions of $E_z$ are vertically oriented for the case of SPPs mode 1, as with other typical SPPs (right panel of Fib. 3b) [49]: PhC mode 2 and SPP mode 2 exhibit similar behaviors in terms of $E_z$ distributions, compared with PhC mode 1 and SPPs mode 1, respectively. The plasmonic and photonic bandgaps (driven by different energy densities between SPPs/PhC mode 1 and 2) are opened at the wavelength of 707 nm (424 THz) and 520 nm (577 THz), respectively; another forbidden region (see grey colored box in Fig. 3a) is also observed (photonic modes). Owing to the use of spherical AuNPs (symmetric structural motifs), the obtainable bandgap width ($\Delta\omega/\omega$) is relatively small (e.g., 0.023 at 686 nm and 0.005 at 520 nm); but, it can be further increased by using anisotropic AuNPs (e.g., Au rice) rather than spherical counterparts [49].

Next, we numerically retrieved the effective impedance ($Z$) of optical metamaterial mirror: as the size of individual AuNP is much smaller compared to the wavelength of interest (herein, the hexagonally close-packed AuNP monolayer can be treated as a homogeneous medium), the applications of both homogenization theory and effective parameter retrieving method (numerical iterative method) using the scattering parameters ($S$ parameters) can be justified [50]. Note that the peak of impedance of optical metamaterial mirror layer was observed near at the forbidden regions such as plasmonic and photonic bandgaps (Fig. 3c); the enhancement of impedance can be achieved at the broadband visible frequency (520 nm ~ 750 nm) by the versatile assembly of 150 nm sized, spherical AuNPs monolayer ($Z$ is ranged from 0.5 to 1.2). The huge enhancement of impedance at 500 nm is due to the interband transition of Au rather than reflection phase control (light penetration and energy storage at the surface of Au). According to the complex reflection coefficient (eqn. 4), the reflection amplitude (Fig. 3d) inversely follows the trend of impedance changes along the wavelength of interest (excepting near at the wavelength of 500 nm). In order to more elucidate the effect of SPPs and PhC modes on obtaining magnetic mirror, we compared the impedance and reflection for 150 nm thick, flat Au layer embedded in 160 nm thick cPVP matrix (electric mirror with phase reversal) and the optical metamaterial mirror (lines with orange and grey colors in Fig. 3c-d indicate the optical metamaterial mirror and flat Au layer, respectively); obviously, a very low impedance (less than 0.1) of a flat Au layer (electric mirror) and the resultant phase reversal ($\varphi = \pi$) of reflecting light results in the almost 100 % reflection over a broadband visible frequency in stark contrast to the high impedance surface of optical metamaterial mirror.

To add analysis in more depth, we characterized additional collective set of optical properties including electric ($|E|$)/magnetic ($|H|$) field distributions and the reflection phase for a flat Au layer and optical metamaterial mirror, as summarized in Fig. 4: these studies were carried out at 511 nm and 707 nm wavelength (respectively at the photonic and plasmonic bandgaps of optical metamaterial mirror). Comparison of a flat Au mirror with AuNP monolayer reveals a far more significant impact of high impedance on the control of both reflection phase and electric field distribution near at the surface of mirror (~ 25 nm from the surface of mirror layer). Whereas the light reflecting off a flat Au layer

shows a standing wave with a minimized electric field near at the surface of mirror layer (maximized electric field at one quarter of wavelength from mirror layer) resulting from in-phase superposition between incident and phase reversed ($\varphi = \pi$) reflecting light, the optical metamaterial mirror can exhibit far more electric field right at the surface of mirror layer (Fig. 4a). This is because the phase shift of the light reflecting off the designed metamaterial mirror can be reconfigured to be much smaller than $\pi$ via SPPs and PhC modes-enabled accumulation of phase (i.e., $\varphi = \pi/6.3$ at wavelength of 707 nm and $\varphi = \pi/7.1$ at wavelength of 511 nm). This reconfigured reflection phase through SPPs and PhC modes of optical metamaterial mirror is well revealed by the shifted sinusoidal envelop of standing wave with respect to field profile above a flat Au mirror. It is also worth noting that overall intensity of standing wave above optical metamaterial mirror is weaker than that above a flat Au mirror; evidencing the stored incident light and accumulated phase into the optical metamaterial mirror. Analyzing magnetic field distribution near at the surface of mirror layer further supports the origin of such metamaterial mirror's behavior. The magnetic field amplitude near at the surface of optical metamaterial mirror is found be weaker than that near at a flat Au layer, as shown in Fig. 4b. As such, the optical metamaterial mirror can harness relatively higher impedance, proper accumulation of phase, and thus more concentrated electric field near at its surface compared with a flat Au counterpart.

*3.3 Flexibility in tuning reflection behavior of AuNP optical metamaterial mirror*

The versatile controllability of optical properties can accrue if we use spherical AuNPs as a building block for the assembly of optical metamaterial mirror rather than conventional lithographic approach to metallic nano-pattern. For example, AuNPs can exist as various structural motifs, such as dielectric core-metallic shell sphere and multilayered metallic alloy sphere [40,51]; also, various sizes of such structural motifs of metallic NPs now can be accessed with recent advances in chemical synthesis [40]. Thereby, through careful tuning of structural features of AuNPs to be assembled into the monolayer, we can explicitly program the properties of SPP or PhC modes (e.g., working wavelength and impedance) and thus reflection behavior of optical metamaterials. Toward this direction, we further expanded the range of available controllability of optical metamaterial properties by using Au shell (15 nm in thickness) and silica core NPs (120 nm in diameter) (abbreviated as silica@Au core-shell NPs).

In the case of silica@Au core-shell NPs monolayer embedded into cPVP matrix (see inset of Fig. 5a), the photonic modes disappear at the wavelength of interest mainly due to the negligible difference of refractive index between silica and cPVP. Thus, the dramatic enhancement of impedance and the resultant reduction of reflection have their origin in SPPs (Fig. 5a-c) modes instead of PhC modes: both SPPs mode 1 (fundamental mode at wavelength of 790 nm) and mode 2 (higher mode at wavelength of 610 nm) simultaneously show the highly concentrated electric field at the vicinity of spherical silica@Au core-shell NPs (Fig. 5c) contrary to photonic modes (see Fig. 4a). Furthermore, the implementation of core/shell architecture into spherical AuNPs provides additional flexibility in the controlling working wavelength of SPP modes: for example, the resonant wavelength of the fundamental SPP mode is shifted to 790 nm from 707 nm by inserting silica core (130 nm in diameter) into spherical AuNPs (150 nm in diameter). Meanwhile, the spherical dielectric core/metallic shell geometry makes the conduction path more tortuous, so as to further increase the impedance, compared with solid spherical AuNP monolayer (e.g., 1.25 at SPPs mode 2 and 1.49 at SPPs mode 1). Moreover, at the wavelength of SPPs modes, silica@Au core-shell NP monolayer exhibits weaker intensity of standing wave above mirror layer compared with solid AuNP counterpart, as an evidence of more accumulated phase and light energy ($\varphi = \pi/7.45$ at wavelength of 610 nm and $\varphi = \pi/7.61$ at wavelength of 790 nm) within mirror layer. Nonetheless, especially at SPPs mode 2, the silica@Au core-shell NPs monolayer shows less favorable spatial distribution of electric field, compared with both SPP and PhC modes of solid AuNP monolayer in terms of deep-subwavelength-scaled optoelectronic device application, because of the highly concentrated electric field mainly at the vicinity of silica@Au core-shell NPs rather than at the surface of mirror layer (see electric field distribution via PhC mode in solid spherical AuNP monolayer, shown in Fig 4a). This result emphasizes the important role of the mode characteristics (i.e., the spatial distribution of electric field) on the design of optical metamaterial mirror to be useful for the enhancement of light absorption in deep-subwavelength-scaled optoelectronic devices. In other words, on-demand control of optical modes of metamaterial mirror as well as the precise control of impedance and reflection phase has to be simultaneously carried out according to the device architectures.

## 4. Optical metamaterial mirror-implemented NFGTM with enhanced light absorption

*4.1 Impact of optical metamaterial mirror on the absorption of the devices*

Encouraged by such compelling advantages of using charge storage layer (assembled AuNPs in a monolayer form) for the rational design of optical metamaterial mirror, we finally profiled impact of the AuNP monolayer mirror on the light absorption properties of the 25 nm thick pentacene-graphene hybrid layer within NFGTM: the numerical FDTD simulation was performed with full device architecture. To calculate absorbed photon fraction selectively within pentacene/graphene hybrid layer, the electric field amplitude was integrated over one period of unit cells. Fig. 6a compares the absorbed photon fraction within the active layer of NFGTM for the different mirror designs such as the absence of metallic mirror (simple cPVP matrix without mirror, grey line), a flat Au mirror (green line), 150 nm sized solid AuNP monolayer (light orange line), and silica@Au core-shell NP monolayer (light blue line).

A simple flat Au mirror can still provide the broadband absorption enhancement of the pentacene-graphene hybrid layer. In our case, the thickness of pentacene and graphene hybrid layer (~ 25 nm) is much smaller than the wavelength of interest (deep-subwavelength-scaled); in contrast to semiconducting materials (e.g., Ge) with extremely strong absorption properties [36], the absorption of pentacene is relatively moderate. Thus, its phase accumulation through round trip propagation or optical resonance within such pentacene-graphene layer should make a negligible contribution to the enhancement of broadband light absorption via a flat Au mirror; this enhancement of broadband light absorption mainly originates from the increased overlap between pentacene layer and increased electric field region (maximized at one quarter of wavelength) resulting from the phase reversed reflection. In particular, as the wavelength is reduced, the absorbed photon fraction becomes more significant mainly due to the interband transition-enabled energy storage and reflection phase accumulation at the surface of Au. This aspect is well reflected by the series of spatial distributions of electric field at different wavelength, as shown in Fig. 6b. We can also observe that the intensity of standing wave above a flat Au mirror layer especially at wavelength of 511 nm is further reduced after the implementation of light absorbing pentacene-graphene layer onto the surface of mirror layer (compare the third panel of Fig. 4a with first panel of Fig. 6b).

Meanwhile, despite of an imperfect magnetic mirror ($\varphi$ is a bit larger than zero, as demonstrated in sections 3.2 and 3.3), the use of the close-packed spherical AuNP monolayer (both solid AuNPs and silica@Au core-shell NPs) as a charge storage layer in NFGTM can result in far more enhancement of the broadband light absorption within pentacene-graphene hybrid layer, compared with both a flat Au mirror and a individually dispersed disc-shaped AuNP (see Fig. 2a). Importantly, the solid AuNP monolayer is found to achieve dramatic enhancement of light absorption at broadband visible frequencies from 550 nm to 730 nm (55 ~ 65 % of incident photon), where pentacene can maximize light absorption and thereby efficiency of charge/carrier generation [39]. Fundamental to this broadband light absorption benefitting from the use of hexagonally close-packed solid AuNP monolayer is detailed in series of electric field spatial distribution (Fig. 6c). We can observe that the irradiated light in normal direction can properly accumulate the phase via PhC (below wavelength of 650 nm) and SPPs modes (above wavelength of 650 nm), so as to achieve the desired electric field distribution in terms of enhancing broadband light absorption in the device. Due to the deep-subwavelength-scaled thickness of active channel layer, the optical resonance and possible photon recirculation within active layer cannot be effectively activated, as mentioned above, in that the broadband light absorption is dominantly in regards to the highly concentrated near-field effect, which is properly controlled to be overlapped with the active layer. This is main reason that PhC modes with the ability to force the electric field to be strongly accumulated at the surface of solid AuNP monolayer can be advantageous over the SPPs modes in terms of broadband light absorption of deep-subwavelength-scaled active layer. Such enhanced light absorption within the active layer and the accumulated electric field preferably at the surface of solid AuNP monolayer also results in the reduced intensity of standing wave above the mirror layer.

In stark contrast to solid AuNP counterpart, it turns out that silica@Au core-shell NPs monolayer strongly absorbs the incident light within each individual particle through SPP whispering-gallery modes at shorter wavelength of interest (e.g., 511 nm) rather than reflecting back, resulting in much poorer performance of light absorption. Also, as mentioned in section 3.3, this optical metamaterial mirror likely accumulates the electric field at the vicinity of silica@Au core-shell NPs rather than at its surface, thereby being less favorable for broadband light absorption in deep-subwavelength-scaled optoelectronic device. However, this result implies that the working wavelength of metamaterial mirror and thus spectral response of the device can be precisely tuned by adjusting the structural motifs (geometry of AuNPs). For example, the translation of dielectric core into AuNPs is found to increase the working wavelength of SPPs mode 1 (fundamental mode). Such high flexibility, which can be achieved in a versatile way with advances in chemical synthesis [40], will be critical for enabling robust application of AuNP-based optical metamaterial mirror to the wide range of deep-subwavelength-scaled optoelectronic devices. Very recently, thin film coating of various active materials including organic dye and perovskite has proven to be efficient for the enhancement of optical cross-section of 2D materials [52,53]; consequently, being capable of on-demand controlling working wavelength with respect to each different spectral responsibility of such active materials should be satisfied in design of optical metamaterial mirror. In line with this, we believe our strategy will provide a versatile and powerful route to the rational manipulation of light-matter interaction within deep-subwavelength-scaled optoelectronic devices.

*4.2 Possible options for the assembly and implementation of optical metamaterial mirrors into the NFGTM*

The key to success in the fabrication of optical metamaterial mirror-implemented NFGTM is to obtain the highly qualified monolayer of solid AuNPs or silica@Au core-shell NPs. Toward this direction, the highly uniform, super-spherical AuNPs or silica@Au core-shell NPs should be accessible. In general, the citrate-mediated reduction of Au chloride, which has been widely used for the synthesis of AuNPs, results in the polygonal shaped, dispersive AuNPs; but, very recently, G.-R. Yi and his colleagues conceived the radical idea for the monocrystalline, super-spherical AuNPs through the elective etching vertices or edges of Au octahedral [40]. Furthermore, the well-established strategy for the selective reduction of Au onto the surface of silica colloidal particles allows us to get an access to the uniformly distributed silica@Au core-shell NPs in a reliable way [54].

Additionally, various methods have been established to organize the large-area monolayer of spherical colloidal nanoparticles, including mechanical rubbing [55], spin coating [56], and controlled dip coating [57], over the last two decades. In general, however, the process yield of such methods has been found to be highly dependent on the substrates (e.g., surface chemistry and roughness); thus, such assembly strategies of AuNP monolayer, in some cases, couldn't be compatible with overall fabrication procedure of NFGTM. This problem of incompatibility between NFGTM fabrication and assembly of AuNP monolayer can be effectively addressed by dry transfer printing with controlled adhesion [58−60]: (i) assembly of AuNP monolayer onto a mother substrate and (ii) transfer printing of it into NFGTM. Finally, the dielectric matrix made of polymer (e.g., cPVP) can be conformally coated by atomic layer deposition (ALD).

## 5. Conclusion

Here, we have suggested theoretical designs of optical metamaterial mirror for achieving the enhancement of broadband light absorption in deep-subwavelength-scaled graphene optoelectronic device, while maintaining nonvolatile photonic memory functionality. By examining numerical simulation, we have revealed that the rationally designed charge storage layer (i.e., spherical AuNP monolayer), having an excellent amenability to the device architecture of NFGTM, can be treated as an effective optical magnetic mirror; furthermore, this optical metamaterial mirror can be achieved in a relatively versatile way in stark contrast to conventional lithographic methods. A more exciting prospect is the versatile access to various structural motifs of metallic NPs benefitting from recent advance in chemical synthesis and the ability to sophisticated engineering of optical properties of metamaterial mirror such as impedance, reflection behavior, and reflection phase. With these advances, a more flexibility in the design of deep-subwavelength-scaled optoelectronic devices should now become available for both rational management of photon flow and the enhancement of device performance.

**Acknowledgements**

This work was fully supported by Samsung Research Funding Center for Samsung Electronics (SRFC-MA1402-09).

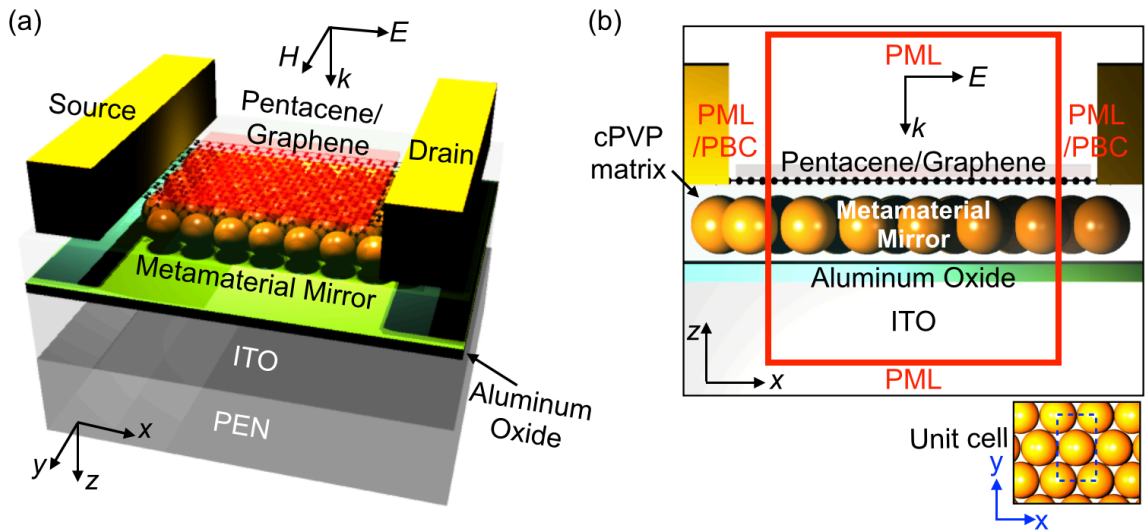

**Figure 1.** (a) Schematic for pentacene-graphene nano-floating gate transistor memory (NFGTM) with optical metamaterial mirror (i.e., hexagonally close-packed, 150 nm-sized spherical AuNP monolayer) (not to scaled). (b) The 3D full field electromagnetic simulation geometry by finite-difference, time-domain (FDTD) method. Perfectly matched layer (PML) and periodic boundary condition (PBC) were employed, as highlighted by red bold line; the blue dotted box indicates the unit cell of FDTD simulation.

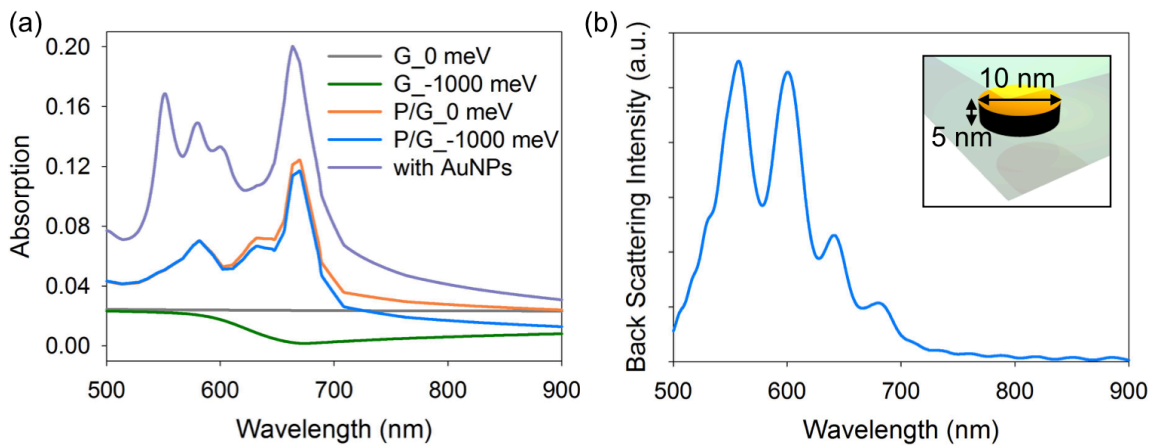

**Figure 2.** (a) Absorption of graphene at charge neutral point (G_0 meV), the gated graphene with -1000 meV of Fermi energy (G_-1000 meV), 25 nm thick pentacene/graphene hybrid with (P/G_-1000 meV) and without gating (P/G_0 meV), and 25 nm thick pentacene/graphene hybrid layer (without gating) with the disc-shaped AuNP (10 nm in diameter and 5 nm in thickness) embedded in cPVP matrix (20 nm thickness, as shown in inset of (b)). (b) Backward scattering intensity of the disc-shaped AuNP (10 nm in diameter and 5 nm in thickness, see inset) embedded in cPVP matrix.

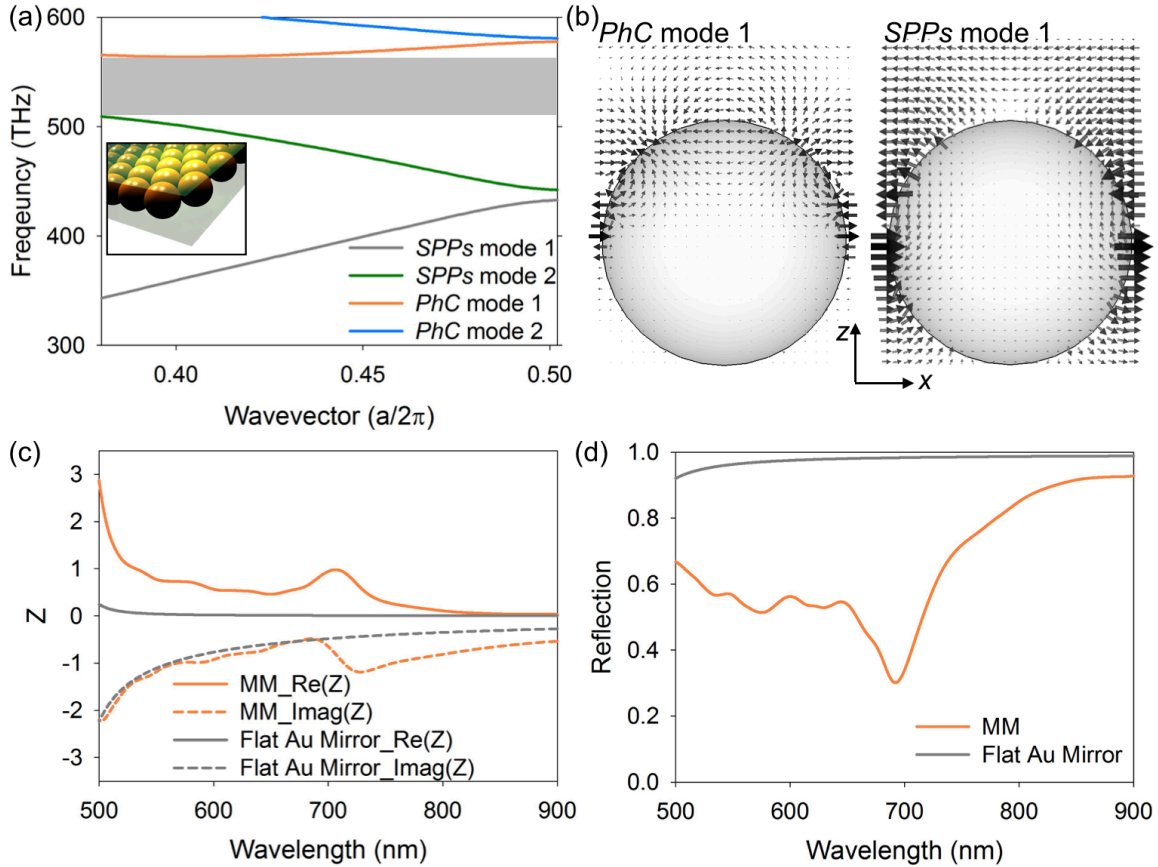

**Figure 3.** Characteristics of the designed optical metamaterial mirror. (a) Calculated band structure of the 150 nm-sized, close-packed spherical AuNP monolayer with hexagonal geometry, which is embedded within 160 nm thick cross-linked poly-4-vinylphenol (cPVP) dielectric matrix (the layer of optical metamaterial mirror, see schematic presented in inset). (b) The spatial distributions of $E_z$ vector for PhC mode 1 (left panel) and SPPs mode 1 (right panel). The length and thickness of black arrow are proportional to the intensity of $E_z$. These were obtained at the plasmonic and photonic bandedges of $k=a/2\pi$. (c) Numerically retrieved impedance of optical metamaterial mirror layer (orange solid and dotted line) and a flat Au layer (grey solid and dotted line). (d) The simulated reflections of optical metamaterial mirror layer (orange bold line) and a flat Au layer (grey bold line).

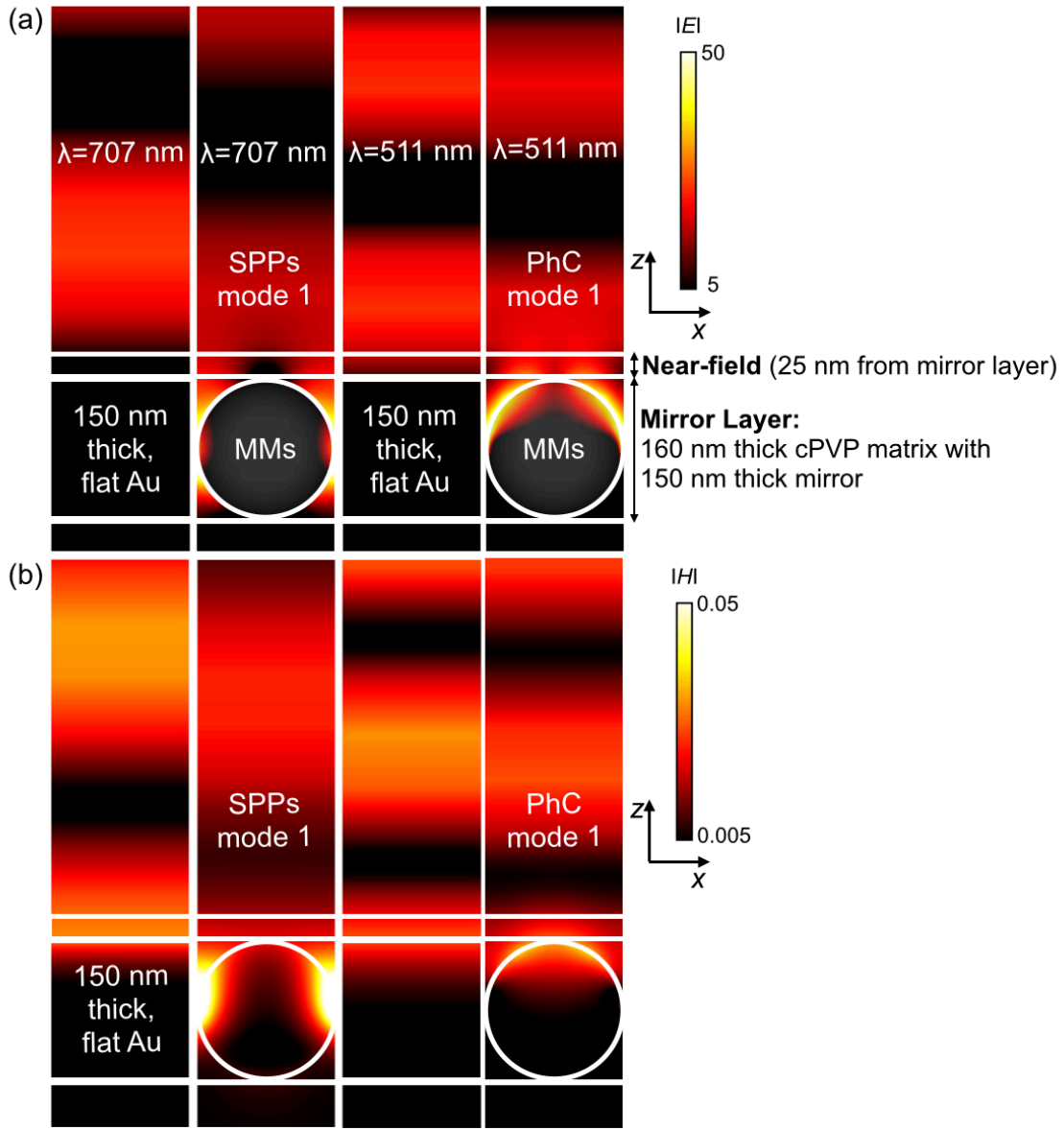

**Figure 4.** (a) |*E*| and (b) |*H*| spatial distributions of a flat Au layer and optical metamaterial layers at wavelength of 707 nm and 511 nm (corresponding to SPPs mode 1 and PhC mode 1 of optical metamaterial mirror, respectively).

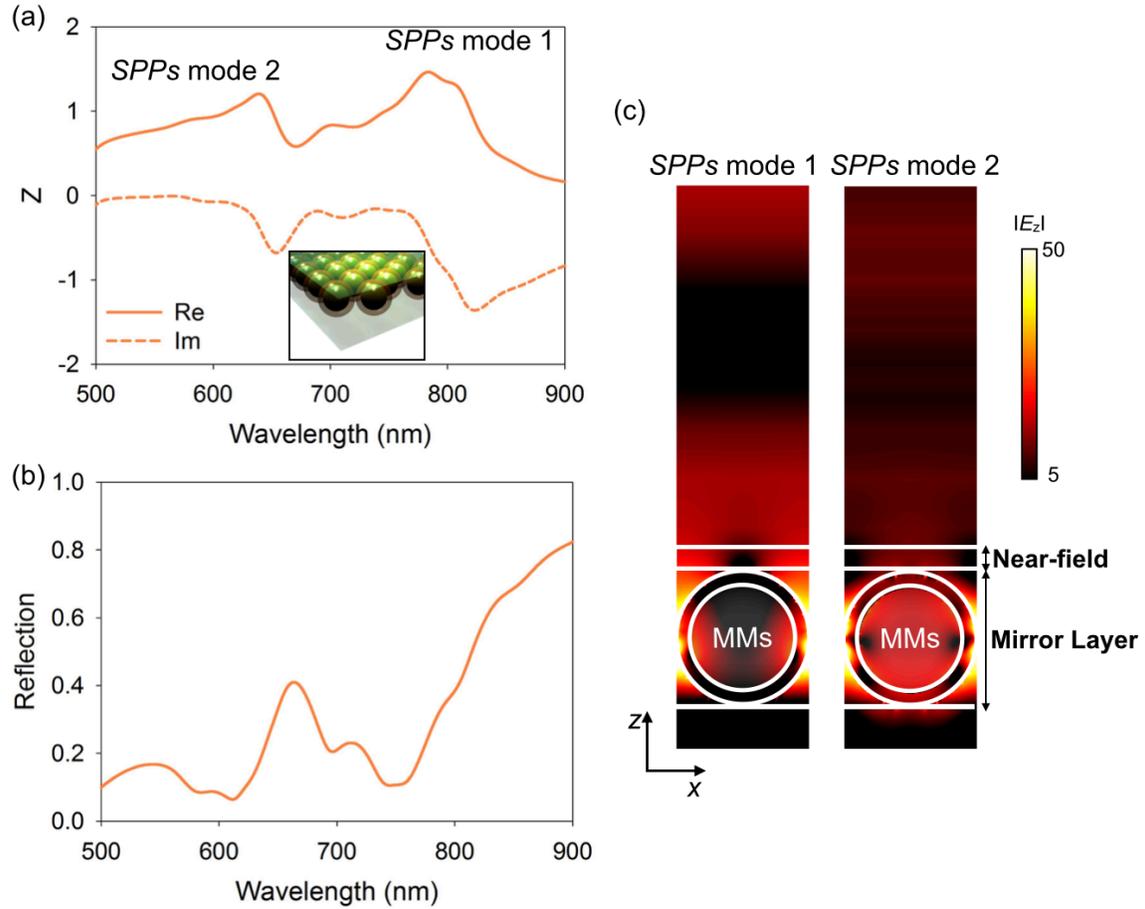

Figure 5. (a) Numerically retrieved impedance and (b) simulated reflection of hexagonally close-packed, silica@Au core-shell NP monolayer, which is embedded into cPVP matrix (160 nm thickness). (c) Electric field distribution of hexagonally close-packed, silica@Au core-shell NP monolayer at SPP mode 1 (left panel) and mode 2 (right panel).

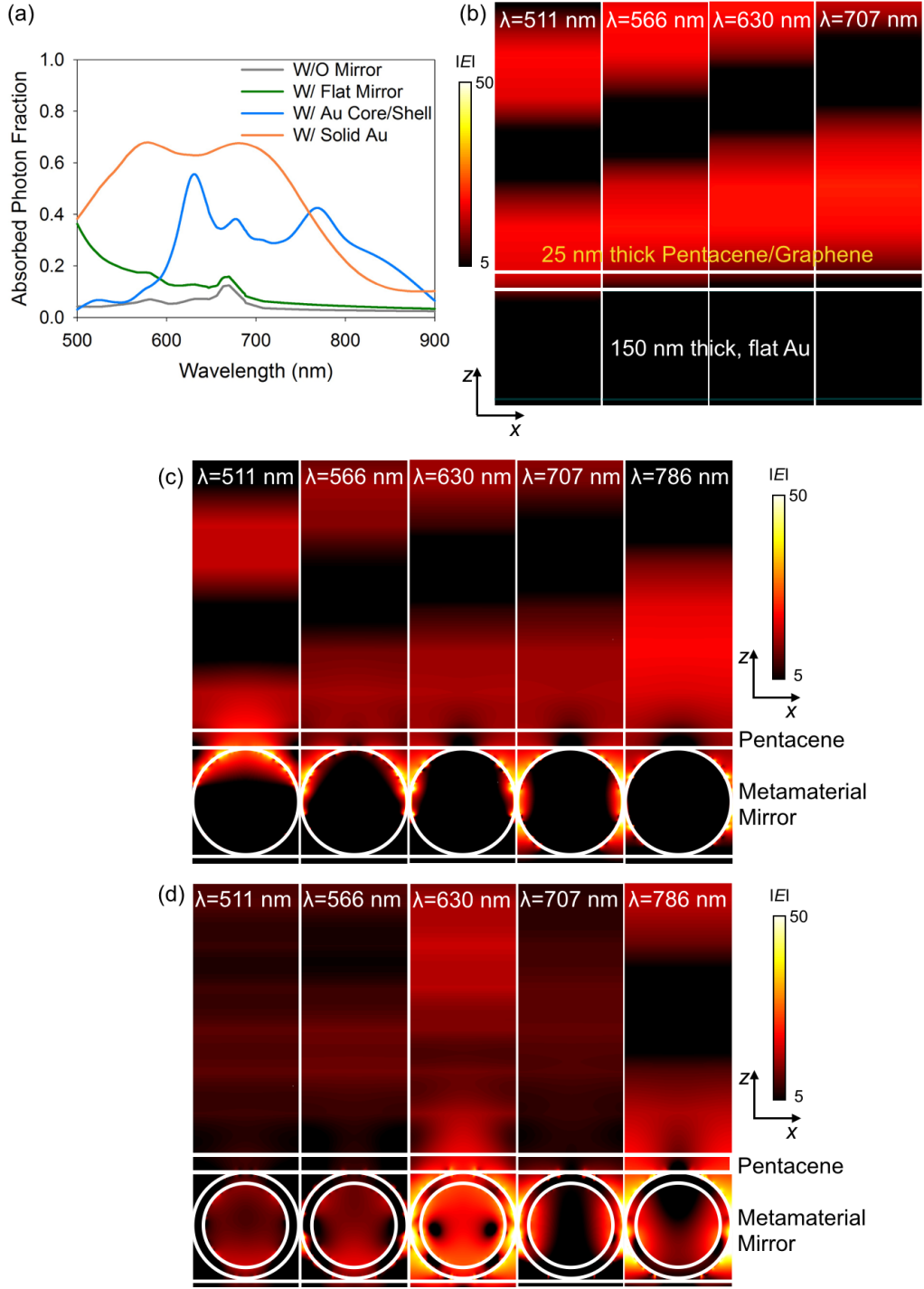

**Figure 6.** (a) Calculated photon absorption fraction within pentacene/graphene hybrid layer for different mirror designs. (b-d) Electric field distribution of the devices with a different mirror layers such as (b) flat Au mirror (thickness of 150 nm), (c) hexagonally close-packed, solid AuNP monolayer, and (d) hexagonally close-packed, silica@Au core-shell NP monolayer.